# Optimized application of double and single layer BEM for in vivo conductivity estimation

Jan C de Munck[1], Andreas Daffertshofer[2], Victoria Montes-Restrepo[3], Theo JC Faes[4], Maureen Clerc[5], Joost Hulshof[6]

[1] Department Physics and Medical Technology, VU University Medical Center, Amsterdam, The Netherlands

[2] Department of Human Movement Sciences, Faculty of Behavioural and Movement Sciences, Vrije Universiteit Amsterdam, Amsterdam Movement Sciences, The Netherlands

[3] Department of Rehabilitation Medicine, Amsterdam Movement Sciences, Amsterdam UMC, Vrije Universiteit Amsterdam, The Netherlands

[4] Department Radiology and Nuclear Medicine, VU University Medical Center, Amsterdam, The Netherlands

[5] Inria Sophia Antipolis, France, and Université Côte d'Azur, France

[6] Department of Mathematics, Beta faculty, Vrije Universiteit Amsterdam, The Netherlands

## Abstract

Inter subject variability of the electrical conductivity of brain, skull and skin strongly limits the accuracy by which current sources underlying electro-encephalography (EEG) can be localized in the brain. This inter subject variability also constrains the possibility to use EEG amplitude parameters as a biomarker to compare the amount of neural activity between different patients. To overcome this problem, one may estimate conductivity parameters in vivo by analyzing the potentials generated by known electric currents, injected into different pairs of EEG electrodes. At present, routine application of this approach is hampered by the computational complexity of the conductivity estimation problem.

Here we analyze the efficiency of this conductivity parameter estimation problem in the context of boundary element method (BEM). We assume geometries of brain, skull and skin compartments are fixed triangular meshes whereas conductivity parameters are treated as unknowns.

We show that a Woodbury update algorithm can be used to obtain a fast conductivity update scheme for both the single and double layer BEM formalism. This algorithm yields a speed gain up to a factor of 20 when compared to the direct computations, apart from at most 50% of additional computation time in the initialization phase of the algorithm. We also derive novel analytically closed expressions for the efficient and accurate computation of BEM matrix elements.

Finally, we discuss which further steps are needed to equip future EEG systems with software devices that enable subject tailored head models for calibrated EEG and accurate source localization, on a routine basis.

## Introduction

The possibility of using multichannel EEG data to localize underlying generators is based on several assumptions translated into a mathematical model that enables the prediction of the observed EEG potentials for a given generator. The source localization problem can hence be formulated as a parameter estimation problem. The accuracy by which EEG source localization is possible depends critically on the realism of the modelling assumptions, like the shape and conductivities of brain, skull and skin, as well as on the geometry and number of sources. The sources are most often modelled as a combination of current dipoles with amplitudes, orientations and/or positions considered as unknowns. Shapes of brain, skull and skin can be extracted from an anatomical MRI scan using image processing techniques [1-3], or they can be obtained by fitting a deformable model to a set of landmarks, like the electrode positions [4,5].

EEG combined with source localization techniques is by far the cheapest and simplest available brain imaging technique and, apart from expensive and immobile MEG, the only technique able to record neural brain activity directly. Clinical applications include pre-surgical mapping of patients with epilepsy, where source localization helps to determine the epileptic focus [6]. Since restoration of brain function after stroke is associated with augmented amplitudes of neuronal currents at the pyramidal cells, EEG has all the ingredients to become an instrument for patient monitoring [7]. In particular, EEG amplitudes measured in a condition where the patient's somatosensory cortex is stimulated by a peripheral external mechanical or electrical stimulus contain valuable information about the recovery of specific brain areas of the patient during rehabilitation. Neural activity at the cerebral cortex is transferred to the EEG sensor by electrical conductance of the tissues between current source and EEG sensor. Hence, applications, source localization and the use of EEG amplitudes as a biomarker for neural activity may be hampered by inaccurate knowledge of electrical conductivity of the tissues between source and sensor. The skull has a particularly strong effect on this signal transfer because its low conductivity dampens the amplitude and smoothens the spatial distribution of the recorded signal [8]. For source localization, this leads to systematic errors in the estimated generators [9-12]. As neural activity biomarker, the observed EEG potentials are confounded by inter subject differences in skull conductivity rendering it virtually impossible to interpret EEG amplitudes as substitutes for neural activity.

Traditionally, the EEG research community has massively used conductivity parameters that were derived from in vitro determination of animal tissue [13] until it was realized that more specific and accurate knowledge on tissue conductivity was required to unlock the unique technological advantages of EEG. New in vitro estimates of skull conductivity became available from freshly excised human brain [14] or skull samples [15]. Some studies differentiate between compact and marrow bone [16]. However, because doubts remained on the relevance of in vitro determinations of tissue conductivity when applied to EEG based inverse modelling, several groups have explored the possibility of in vivo measurement of these parameters, using calibrated AC currents that are injected in epilepsy patients, through intracranial electrodes and recorded with scalp electrodes [17,18]. Also, non-invasive measurement setups have been proposed, where AC currents are injected into scalp electrodes and where the resulting potential differences at the remaining EEG electrodes are analysed [19-23]. This approach, inspired by electrical impedance tomography (EIT), has as the great advantage that tissue conductivities are derived from the individual subject and that physiological changes in excised tissue sample during measurement can be avoided.

With the EIT approach, the head is modelled as a set of compartments of which the geometries are known from MR images and of which only the conductivities are treated as unknown parameters to be determined from the EIT data, i.e. from the potentials resulting from the injected currents. Apart from [23], who implemented a multi compartment model, most often, a three-compartment model is adopted, consisting of brain, skull and skin, wherein the conductivities of skin and brain are assumed to be equal, resulting in a non-linear two parameter estimation problem. In a recent study, [24] showed that only a short epoch of EIT data is sufficient for in vivo skull determination on an individual subject (in total only several minutes of data suffice). This result implies the feasibility to apply the EIT approach on a routine basis, when the switching of injected currents over different electrode pairs is automated. In this way, it may become standard practice to setup calibrated head models for EEG that are tailored to the individual subject. However, since the conductivity estimation problem is a non-linear parameter estimation problem, iterative techniques are unavoidable whereby in each iteration the potential distribution due the injected current needs to be computed, which is computationally expensive for realistic models. Therefore, at present routine application of EIT remains hampered by the computational aspects of the EIT data analysis problem.

Analytical solutions of the EIT forward problem, i.e. to predict the potential distribution from known current injections, are only available for models wherein the head is described as a set of concentric spheres [25]. Such models, however, lack geometrical accuracy and likely lead to unsatisfactory results. Realistic geometries have been modelled in the context of EIT using the finite element method (FEM) [23] as well as the boundary element method (BEM) [20-22]. The computational challenge of conductivity fitting is that the forward problem needs to be solved for many different combinations of conductivities. For the simplest BEM variant, the double layer collocation approach, a computationally efficient update of the EIT forward problem has been proposed by [23]. Other approaches require complete new forward calculations, for each new set of conductivity parameters. Therefore, we expect that the BEM has most potential to serve as a computational tool in routine applications of EIT to generate calibrated head models and in this paper several BEM variants are explored in detail. A general approach to find fast approximate solutions of the EIT forward problem is presented in [26].

We here present the single layer and double layer BEM in a common mathematical framework and compare these two approaches in the context of the estimation of conductivity parameters. Specifically, formulas are derived wherein substantial speed gain can be achieved when using Woodbury updates for both single layer and double layer BEM, using the Galerkin as opposed to the collocation method to discretize the boundary integral equations. The expressions derived in [27] for computation of BEM matrix elements are generalized to the single layer BEM. Finally, four different BEM variants (single and double layer, piece constant and piecewise linear interpolation) are compared in terms of speed and accuracy. Preliminary results of this work have been presented as conference proceedings [28,29]. C++ source codes of the presented methods are available through https://github.com/jcdemunck5.

## Methods

*Definition of symbols and problem*

With the BEM the conductor is described as an object consisting of compartments with different conductivities. For simplicity it is here assumed that the compartments are bounded by $K$ closed nested surfaces $\Gamma_0, \Gamma_1, \Gamma_2,.., \Gamma_{K-1}$, although BEM is also applicable to cases where the surfaces are not closed or nested [30,31]. The conductivity parameters are indicated by $\sigma_k^+$ and $\sigma_k^-$, and represent the conductivities of the compartments just out- and inside surface $\Gamma_k$. In the context of this paper $\Gamma_0$ is the outer surface where the currents are injected. These currents are represented by a known function $j_0(x)$, with

$$\oiint_{\Gamma_0} j_0(x) \mathrm{d}S_x = 0 \qquad (1)$$

In practice, $j_0(x)$ will be zero everywhere, except at two small finite regions where the current is injected and extracted. The resulting potential distribution $\psi(x)$ is the solution of the following system of partial differential equations

$$\begin{cases} \Delta\psi = 0 & x \in \Omega_0 \cup \Omega_1 \cup \ldots \cup \Omega_{L-1} \\ \psi^+ = \psi^- & x \in \Gamma_0 \cup \Gamma_1 \cup \ldots \cup \Gamma_{K-1} \\ \sigma^+ \partial_n \psi^+ = \sigma^- \partial_n \psi^- & x \in \Gamma_0 \cup \Gamma_1 \cup \ldots \cup \Gamma_{K-1} \\ \sigma^- \partial_n \psi^- = j_0(x) & x \in \Gamma_0 \end{cases} \quad (2)$$

Here $\Omega_l$, $l = 0, \ldots L - 1$ are the open regions in which the conductivity is constant and $\partial_n \psi$ denotes the derivative of $\psi$ in the direction of the surface normal. That is, the injected current acts as a Neumann boundary condition. The main symbols are illustrated in Figure 1.

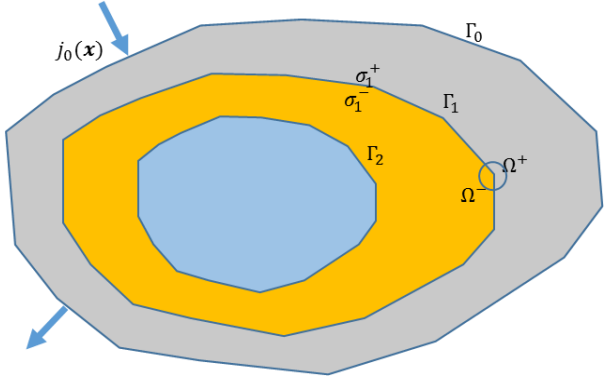

*Figure 1. Definition of the main symbols. The volume conductor consists of a set of nested compartments with different conductivity. In the idealized conductor the compartment interfaces $\Gamma_k$ are smooth surfaces but after discretization they become piecewise linear polyhedrons. The boundary integral equations used in this paper are valid for these polyhedron surfaces and they may therefore contain sharp corners, in this figure indicated by $\Omega^-$ and $\Omega^+$.*

### BEM integral equations

The BEM is based on the reformulation of (2) in terms of integral equations, where the unknown $\psi$ is confined to the compartment boundaries. These equations can be solved by discretizing the integrals on triangular mesh representations of these boundaries. Because the triangulated surfaces are piecewise linear, they will contain sharp corners and ridges. Contrary to other studies, here the presence of these corners is accounted for when setting up the BEM integral equations. There are different approaches to convert (2) to boundary integral equations. In a seminal study by [32] three of them are treated in detail for the dipole source case: the single and double layer formalism and the symmetric BEM. The latter gave most favorable results for dipole sources. However, here we focus on the first two approaches because it appears that these variants allow efficient conductivity updates when the BEM is used for conductivity estimation. For the symmetric BEM a similar update scheme does not seems to be possible.

To find boundary integral forms of (2) the following operators are used

$$\left(\hat{S}_{ik}\psi_k\right)(y) = \lim_{\varepsilon \to 0} \oiint_{\Gamma_k \setminus \partial B(y,\varepsilon)} \psi_k(x) G(x - y) dS_y \qquad \text{with } y \in \Gamma_i \qquad (3)$$

$$\left(\widehat{D}_{ik}\psi_k\right)(y) = \lim_{\varepsilon \to 0} \oiint_{\Gamma_k \setminus \partial B(y,\varepsilon)} \psi_k(x) \nabla_x G(x - y) \cdot n(x) dS_x \qquad \text{with } y \in \Gamma_i \qquad (4)$$

$$\left(\widehat{D}^*_{ik}\psi_k\right)(y) = \lim_{\varepsilon \to 0} \oiint_{\Gamma_k \setminus \partial B(y,\varepsilon)} \psi_k(x) \nabla_y G(x - y) \cdot n(y) dS_x \qquad \text{with } y \in \Gamma_i \qquad (5)$$

In (3-5), $G$ is the fundamental solution of the Laplace equation,

$$G(x - y) \equiv \frac{1}{4\pi |x-y|} \qquad (6)$$

i.e. $\Delta G(x - y) = -\delta(x - y)$. The operators $\hat{S}_{ik}$, $\widehat{D}_{ik}$ and $\widehat{D}^*_{ik}$ map functions defined on $\Gamma_k$ to functions defined on $\Gamma_i$. When $i = k$ there are singularity issues when the integration variable $x$ approaches $y$. In the integration, a small sphere around $y$ of radius $\varepsilon$ is excluded from $\Gamma_k$ and replaced by a spherical cap, and then the limit $\varepsilon \to 0$ is taken. In the sequel the combination of surface integral and taking the limit is indicated by $\oiint^*$. Equation (5), known as the adjoint double layer integral, can be obtained by taking the $y$ gradient of (3). Contrary to the double layer integral (4), the normal $n()$ refers to the local normal at $y$ and does depend on the integration variable. For non-smooth surfaces, this comes with the difficulty that when $y$ is on a sharp corner, the local normal is not uniquely defined. This difficulty can be overcome by taking the Galerkin approach to discretize the integral equations that are equivalent to (2).

When the integrals in (3) to (5) are used to generate potential distributions outside the surfaces $\Gamma_i$, one has to take care that either this potential, or its normal derivative makes a jump when $y$ crosses $\Gamma_i$. In particular, when $y$ crosses a non-smooth part of the surface, this jump depends on the local solid angle of the surface, subtended at the crossing point. In Appendix A we sketch how these properties can serve to derive the following familiar boundary integral equations. More mathematical details are presented in textbooks, such as [33,34]. One finds

$$\frac{\sigma_i^+ \Omega_i^+ + \sigma_i^- \Omega_i^-}{4\pi} \psi_i - \sum_k (\sigma_k^+ - \sigma_k^-) \widehat{D}_{ik} \psi_k = \hat{S}_{i0} j_0 \tag{7}$$

for the double layer formalism, and

$$\begin{cases} \frac{\sigma_i^+ \Omega_i^+ + \sigma_i^- \Omega_i^-}{4\pi} \varphi_i - (\sigma_i^+ - \sigma_i^-) \sum_k \widehat{D}_{ik}^* \varphi_k = \delta_{i0} j_0 \\ \psi_i = \sum_k \hat{S}_{ik} \varphi_k \end{cases} \tag{8}$$

for the single layer formalism. There, $\Omega_i^+(x)$ and $\Omega_i^-(x)$ are the inner and outer solid angles by which surface $\Gamma_i$ is viewed at $x$ and $\delta_{i0}$ is the Kronecker delta. Put differently, with the single layer formalism the potential is computed in two steps whereas the double layer formalism gives the potential directly.

*Discretization*

The potential $\psi_k(x)$ on $\Gamma_k$ is discretized using a set of interpolation functions $h_k^n(x)$ that are derived from a triangular mesh approximating $\Gamma_k$

$$\psi_k(x) \approx \sum_{n=0}^{N_k-1} \psi_k^n h_k^n(x) \qquad \text{for } x \in \Gamma_k \tag{9}$$

The nodes of the triangular mesh are represented by $x_i^m$, where the lower indices represent surface number and the upper indices the triangles or node points on that surface. Similarly, we approximate $j_k(x)$ and $\varphi_k(x)$

$$\begin{cases} j_k(x) \approx \sum_{n=0}^{N_k-1} j_k^n h_k^n(x) \\ \varphi_k(x) \approx \sum_{n=0}^{N_k-1} \varphi_k^n h_k^n(x) \end{cases} \tag{10}$$

In the collocation approach of BEM, the values of $x$ are varied over the node points in order to discretize the continuous integral equations. As said, we focus on the Galerkin approach [35], where discretization of the integral equations is accomplished by multiplication with $h_i^m(x)$ and integrating over $\Gamma_i$. In the sequel, this operation is notated as

$$\langle \xi, h_i^m \rangle_i \equiv \oiint^*_{\Gamma_i} \xi(y)\, h_i^m(y)\, dS_y \tag{11}$$

The lower index of the inner product $\langle , \rangle_i$ is dropped when it is clear from the context over which surface the integration occurs. The main advantage of the Galerkin approach is that integral equations (7) and (8) are valid in the weak formulation, meaning that when the difference of left and right hand side is multiplied with a class of test functions and integrated, one obtains identically zero for all test functions from that class.

Using $\frac{\sigma_i^+ \Omega_i^+ + \sigma_i^- \Omega_i^-}{4\pi} = \sigma_i^+ - (\sigma_i^+ - \sigma_i^-) \frac{\Omega_i^-}{4\pi}$, one finds for the double layer BEM that

$$\sigma_i^+ \sum_n \langle h_i^m, h_i^n \rangle \psi_i^n - \sum_{k,n}(\sigma_k^+ - \sigma_k^-)\langle h_i^m, (\widehat{D}_{i,k} + \tfrac{\Omega_k^-}{4\pi}\delta_{i,k})h_k^n \rangle \psi_k^n = \sum_n \langle h_i^m, \hat{S}_{i,0} h_0^n \rangle j_0^n \quad (12)$$

Using that $\widehat{D}_{ik}^*$ and $\widehat{D}_{i,k}$ are adjoint integral operators, i.e. that $\langle \zeta, \widehat{D}_{ik}^*\xi \rangle_i = \langle \widehat{D}_{k,i}\zeta, \xi \rangle_k$ where $\xi$ and $\zeta$ are functions on $\Gamma_k$ and $\Gamma_i$ respectively, one similarly finds for the single layer BEM

$$\begin{cases} \sigma_i^+ \sum_n \langle h_i^m, h_i^n \rangle \varphi_i^n - (\sigma_i^+ - \sigma_i^-) \sum_{k,n}\langle (\widehat{D}_{k,i} + \tfrac{\Omega_k^-}{4\pi}\delta_{k,i})h_i^m, h_k^n \rangle \varphi_k^n = \sum_n \langle h_i^m, h_0^n \rangle j_0^n \\ \sum_n \langle h_i^m, h_i^n \rangle \psi_i^n = \sum_{k,n}\langle h_i^m, \hat{S}_{i,k} h_k^n \rangle \varphi_k^n \end{cases} \quad (13)$$

Note that in this way the adjoint double layer operator $\widehat{D}_{ik}^*$ has been eliminated from the equations. By this, ambiguities with the normal $\boldsymbol{n}$ at sharp corners are confined to integration regions of zero measure.

Next, we introduce the following block matrices where lower indices refer to surfaces

$$(G_{i,k})^{m,n} = \langle h_i^m, (\widehat{D}_{k,i} + \tfrac{\Omega_k^-}{4\pi}\delta_{k,i})h_k^n \rangle \quad (14)$$

$$(H_{i,k})^{m,n} = \langle h_i^m, h_i^n \rangle \delta_{k,i} \quad (15)$$

$$(S_{i,k})^{m,n} = \langle h_i^m, \hat{S}_{i,k} h_k^n \rangle \quad (16)$$

When piecewise constant interpolation is used, $H$ is a diagonal matrix with the triangle areas on the diagonal. For piecewise linear interpolation, the functions $h_i^m(\boldsymbol{y})$ overlap for neighboring elements, which results in a sparse symmetric matrix (with stable inverse). We note that $S$ is symmetric while $G$ is not. Computational details on the analytical and numerical integration of these matrix elements are given in Appendix B.

From here on, vectors of variables on a certain surface are represented in bold face. We find for the double layer formalism

$$\sigma_i^+ H_{i,i}\boldsymbol{\psi}_i - \sum_k (\sigma_k^+ - \sigma_k^-) G_{i,k}\boldsymbol{\psi}_k = S_{i,0}\boldsymbol{j}_0 \quad (17)$$

Similarly, for the single layer formalism one finds

$$\begin{cases} \sigma_i^+ H_{i,i}\boldsymbol{\varphi}_i - (\sigma_i^+ - \sigma_i^-) \sum_k G_{k,i}^T \boldsymbol{\varphi}_k = H_{i,0}\boldsymbol{j}_0 \\ H_{i,i}\boldsymbol{\psi}_i = \sum_k S_{i,k}\boldsymbol{\varphi}_k \end{cases} \quad (18)$$

In (18), $T$ is used to indicate the transpose of a matrix or vector. The Neumann problem (2) is singular (the solution is determined up to a constant) and this is reflected in the singularity of the integral equations (7) and (8). In Appendix C, as well as in [36] it is shown that, due to some properties of the interpolation functions, the singularity of the integral equations is inherited by the discretized systems, i.e. it has a right eigenvector $\boldsymbol{e} \equiv (1, \ldots, 1)^T$ corresponding to eigenvalue zero. To remove this singularity, we define the blocked matrix $A_\sigma$ as follows

$$(A_\sigma)_{i,k} \equiv \sigma_i^+ H_{i,i}\delta_{i,k} - (\sigma_k^+ - \sigma_k^-)G_{i,k} + \lambda \boldsymbol{e}_i \boldsymbol{e}_k^T \quad (19)$$

Note that the zero eigenvalue has been removed by adding a matrix of ones, multiplied with an arbitrary scaling constant $\lambda$, usually $1/N$, where $N$ is the total number of unknowns. The vector $\boldsymbol{e}_k$ is a column vector with $N_k$ ones. By this addition the solution of the discretized system satisfies the constraint that the sum of the potentials is zero. Now, if all variables corresponding to one surface are combined in one vector (e.g. $\boldsymbol{\psi} = (\boldsymbol{\psi}_0^T, \boldsymbol{\psi}_1^T, \ldots, \boldsymbol{\psi}_{K-1}^T)^T$) equations (17) and (18), augmented with the constraint $\boldsymbol{e}^T \boldsymbol{\psi} = 0$, can be compactly represented as

$$\begin{cases} A_\sigma \boldsymbol{\psi} = S\boldsymbol{j} \\ A_\sigma^T \boldsymbol{\varphi} = H\boldsymbol{j} \text{ and } \boldsymbol{\psi} = H^{-1}S\boldsymbol{\varphi} \end{cases} \quad (20)$$

In other words, in the double layer case one has $\boldsymbol{\psi} = A_\sigma^{-1} S\boldsymbol{j}$ and in the single layer case $\boldsymbol{\psi} = H^{-1}SA_\sigma^{-T}H\boldsymbol{j}$. For the single layer case the transposed system is solved as for the double layer case and the order in which $S$ is applied is reversed. The matrix $H$ and its inverse explicitly occur in the single layer case, and not in the double layer case. We note that in practical situations we are only interested in the potential at the outer surface, so that in the single layer case one only has to consider a small subset of the rows of $S$ to get the potentials at the EEG electrodes.

*Conductivity updates*

In our specific application, the conductivity of the brain, skull and skin are estimated from potentials that are generated by known injected currents. Generally, a nonlinear fitting algorithm is used where the differences between observed and theoretical EIT potentials are minimized, which requires model predictions for many different combinations of brain, skull and skin conductivities. We demonstrate how these predictions can be computed efficiently making use of the special structure of the BEM system matrix $A_\sigma$

For simplicity, we assume that there are three nested surfaces $\Gamma_0$ (outer), $\Gamma_1$ (middle) and $\Gamma_2$ (inner), with $\sigma_0 \equiv \sigma_0^- = \sigma_1^+$, $\sigma_1 \equiv \sigma_1^- = \sigma_2^+$ and $\sigma_2 \equiv \sigma_2^-$. In our application $\sigma_0$, $\sigma_1$ and $\sigma_2$ represent the conductivities of skin, skull and brain, respectively, and we have $\sigma_0^+ = 0$. With the following notation

$$L_\sigma \equiv \begin{pmatrix} \sigma_0 I_{N_0} & 0 & 0 \\ 0 & (\sigma_1 - \sigma_0)I_{N_1} & 0 \\ 0 & 0 & (\sigma_2 - \sigma_1)I_{N_2} \end{pmatrix} \tag{21}$$

and

$$\Lambda_\sigma \equiv \begin{pmatrix} 0 & 0 & 0 \\ 0 & \sigma_0 H_{11} & 0 \\ 0 & 0 & \sigma_1 H_{22} \end{pmatrix} \tag{22}$$

we have to consider the following system matrix for the double layer formalism (i.e. $A_\sigma \psi = S\boldsymbol{j}$ )

$$A_\sigma = GL_\sigma + \Lambda_\sigma + \lambda\, \boldsymbol{e}\boldsymbol{e}^T \tag{23}$$

Note that the zeroes in (21) and (22) represent sub matrices (blocks) filled with zeroes. The only dependency on the conductivity parameters occurs in the diagonal matrix $L_\sigma$ and in the matrix $\Lambda_\sigma$, which is sparse and has a large zero block in the upper left corner. The size of this zero-block corresponds to the number of unknowns at the outer surface. Because largest potential gradients occur near the injection electrodes, an efficient implementation of the BEM allocates relatively many unknowns at $\Gamma_0$ [20] and therefore the zero-block will be typically sized $\frac{1}{2}N \times \frac{1}{2}N$.

Let $A_{old}$ be the system matrix for some starting values of an iterative algorithm to fit the conductivities, e.g., $\sigma_0 = 1$, $\sigma_1 = 2$ and $\sigma_0 = 3$. Then, one way to update the matrix $A_\sigma$ for new values of $\boldsymbol{\sigma}$ is to eliminate $G$ from (23),

$$G = (A_\sigma - \Lambda_\sigma - \lambda \boldsymbol{e}\boldsymbol{e}^T)L_\sigma^{-1} \tag{24}$$

and compute $A_\sigma$ as

$$A_\sigma = (A_{old} - \Lambda_{old} - \lambda \boldsymbol{e}\boldsymbol{e}^T)L_{old}^{-1}L_\sigma + \Lambda_\sigma + \lambda\, \boldsymbol{e}\boldsymbol{e}^T \tag{25}$$

This will save the re-computation of $G$ but still requires a full $N \times N$ LU decomposition of $A_\sigma$ for every new of $\boldsymbol{\sigma}$. Hence, we express $A_\sigma$ as

$$A_\sigma = GL_\sigma + UV_\sigma^T \tag{26}$$

with

$$U \equiv \begin{pmatrix} 0 & 0 & \lambda \boldsymbol{e}_{N_0} \\ H_{11} & 0 & \lambda \boldsymbol{e}_{N_1} \\ 0 & H_{22} & \lambda \boldsymbol{e}_{N_2} \end{pmatrix} \quad \text{and} \quad V_\sigma \equiv \begin{pmatrix} 0 & 0 & \boldsymbol{e}_{N_0} \\ \sigma_0 I_{N_1} & 0 & \boldsymbol{e}_{N_1} \\ 0 & \sigma_1 I_{N_2} & \boldsymbol{e}_{N_2} \end{pmatrix} \tag{27}$$

If one has computed the solution $\boldsymbol{\psi}_{old}$ of the system

$$A_{old}\boldsymbol{\psi}_{old} = S\boldsymbol{j} \tag{28}$$

for fixed conductivities and one needs to find the solution $\boldsymbol{\psi}_\sigma$ of

$$A_\sigma \boldsymbol{\psi}_\sigma = S\boldsymbol{j} \tag{29}$$

then one can exploit that $G = (A_{old} - UV_{old}^T)L_{old}^{-1}$ in order to find

$$A_\sigma = (A_{old} - UV_{old}^T)L_{old}^{-1}L_\sigma + UV_\sigma^T = (A_{old}L_{old}^{-1} + UW_\sigma^T)L_\sigma \tag{30}$$

with $W_\sigma^T \equiv V_\sigma^T L_\sigma^{-1} - V_{old}^T L_{old}^{-1}$. In more detail, one has to consider

$$W_\sigma^T = \begin{pmatrix} 0 & \left(\frac{\sigma_0}{\sigma_1-\sigma_0} - \frac{\sigma_0^{old}}{\sigma_1^{old}-\sigma_0^{old}}\right)I_{N_1} & 0 \\ 0 & 0 & \left(\frac{\sigma_1}{\sigma_2-\sigma_1} - \frac{\sigma_1^{old}}{\sigma_2^{old}-\sigma_1^{old}}\right)I_{N_2} \\ \left(\frac{1}{\sigma_0} - \frac{1}{\sigma_0^{old}}\right)\boldsymbol{e}_{N_0}^T & \left(\frac{1}{\sigma_1-\sigma_0} - \frac{1}{\sigma_1^{old}-\sigma_0^{old}}\right)\boldsymbol{e}_{N_1}^T & \left(\frac{1}{\sigma_2-\sigma_1} - \frac{1}{\sigma_2^{old}-\sigma_1^{old}}\right)\boldsymbol{e}_{N_2}^T \end{pmatrix} \tag{31}$$

The ranks of $U$ and $W_\sigma$ are $N_1 + N_2 + 1$. In consequence, (30) implies that apart from the sparse matrices $L_{old}$ and $L_\sigma$, the difference between $A_\sigma$ and $A_{old}$ is a rank deficient matrix. In this case, the Woodbury matrix identity [37] can be applied to find a solution of (29) provided the solution of (28) is given. According to the Woodbury approach for any set of matrices $A$, $U$ and $V$ for which $A^{-1}$ and $(A + UV^T)^{-1}$ exist one has

$$(A + UV^T)^{-1} = A^{-1} - A^{-1}U(I + V^T A^{-1}U)V^T A^{-1} \tag{32}$$

Applied to our situation, equation (30), with $A = A_{old}L_{old}^{-1}L_\sigma$ and $V = L_\sigma W_\sigma$, we find the inverse of $A_\sigma$ in terms of

$$A_\sigma^{-1} = L_\sigma^{-1} L_{old} A_{old}^{-1} - L_\sigma^{-1} L_{old} A_{old}^{-1} U (I_{N_1+N_2+1} + W_\sigma^T L_{old} A_{old}^{-1} U)^{-1} W_\sigma^T L_{old} A_{old}^{-1} \tag{33}$$

and the following update rule for $\boldsymbol{\psi}_\sigma$

$$\boldsymbol{\psi}_\sigma = L_\sigma^{-1}\left(I_N - L_{old}A_{old}^{-1}U(I_{N_1+N_2+1} + W_\sigma^T L_{old} A_{old}^{-1} U)^{-1} W_\sigma^T\right) L_{old}\boldsymbol{\psi}_{old} \tag{34}$$

When introducing $Y_{old}$ by means of

$$Y_{old} = L_{old}A_{old}^{-1}U \tag{35}$$

this can be expressed concisely as

$$\boldsymbol{\psi}_\sigma = L_\sigma^{-1}\left(I_N - Y_{old}(I_{N_1+N_2+1} + W_\sigma^T Y_{old})^{-1} W_\sigma^T\right) L_{old}\boldsymbol{\psi}_{old} \tag{36}$$

Please note that $Y_{old}$ depends exclusively on the "old" conductivities and therefore needs to be computed only once. For every new combination of $\boldsymbol{\sigma} = (\sigma_0, \sigma_1, \sigma_2)^T$, one can solve $\boldsymbol{x}_\sigma$ using

$$(I_{N_1+N_2+1} + W_\sigma^T Y_{old})\boldsymbol{x}_\sigma = W_\sigma^T L_{old}\boldsymbol{\psi}_{old} \tag{37}$$

and compute $\boldsymbol{\psi}_\sigma$ via

$$\boldsymbol{\psi}_\sigma = L_\sigma^{-1} L_{old}(\boldsymbol{\psi}_{old} - Y_{old}\boldsymbol{x}_\sigma) \tag{38}$$

The important advantage is that instead of an LU-decomposition of a full *N*×*N* matrix, only the LU decomposition of an $\frac{1}{2}N \times \frac{1}{2}N$ matrix is needed in (37).

For the single layer formalism an analogous approach is possible. One has to find $\boldsymbol{\varphi}_\sigma = A_\sigma^{-T}H\boldsymbol{j}$ and the application of the Woodbury matrix identity yields

$$\boldsymbol{\varphi}_\sigma = A_{old}^{-T} L_{old}\left(I_N - W_\sigma(I_{N_1+N_2+1} + W_\sigma^T Y_{old})^{-T} Y_{old}^T\right) L_\sigma^{-1} H\boldsymbol{j} \tag{39}$$

One also precomputes $Y_{old}$ and the LU decomposition of $A_{old}$. For each new $\boldsymbol{\sigma}$ one solves $\boldsymbol{z}_\sigma$ from

$$\left(I_{N_1+N_2+1} + W_{\sigma}^T Y_{old}\right)^T \boldsymbol{z}_{\sigma} = Y_{old}^T L_{\sigma}^{-1} H \boldsymbol{j} \tag{40}$$

and finds the single layer density by solving

$$A_{old}^T \boldsymbol{\varphi}_{\sigma} = L_{old} L_{\sigma}^{-1} H \boldsymbol{j} - W_{\sigma} \boldsymbol{z}_{\sigma} \tag{41}$$

With the single layer formalism matrix multiplications are performed in the opposite order as in the double layer formalism. Hence, we end up with the application of the inverse of $A_{old}^T$ and have to keep its LU decomposition in memory. Finally, the potentials at the electrodes can be computed using

$$\boldsymbol{\psi}_{\sigma} = H^{-1} S \boldsymbol{\varphi}_{\sigma} \tag{42}$$

We note that the standard choice for the scaling parameter $\lambda = \frac{1}{N}$ does not suffice for the single layer formalism. We chose $\lambda = {\sigma_1}/{(\sigma_0 + \sigma_2)}$ to avoid numerical instabilities and to get exact agreement between the Woodbury and the direct computations.

*Optimized algorithm and complexity*

The different steps of the algorithm to compute $\boldsymbol{\psi}_{\sigma}$ using the Woodbury approach are summarized in Table 1 for the double layer formalism. The last column in that table gives estimates of the computational complexity of each step. In the preparation phase, $\boldsymbol{\sigma}_{old}$ is chosen (which can be done arbitrarily, as long as neighbouring compartments have different conductivity) and current injections used to acquire the EIT data are stored in $N_{inj}$ vectors $\boldsymbol{j}$. Typically, the number of injections is small (10 to 60). The complexity of the computation of the matrix elements of $A_{old}$ and $S$ is $\mathcal{O}(N^2)$, but the precise factor strongly depends on the implementation of the numerical integration. Since the matrix $U$ is dimensioned $N \times \frac{1}{2}N$, the computation of $Y_{old}$ in step 4 of Table 1 requires $\frac{1}{2}N$ times the application of the LU decomposition of $A_{old}$, assuming that half of the BEM nodes are located on the outer surface, which amounts to $\mathcal{O}(\frac{1}{2}N^3)$ operations. For every new value of $\boldsymbol{\sigma}$, requested by the conductivity fit algorithm, step 6 has to be executed once, followed by steps 7 and 8 once for every current injection. Step 7 is dominated by the application of the LU decomposition determined in step 6. Finally, the matrix vector multiplications performed in step 8 are confined to the rows corresponding to the $N_{elc}$ EEG electrodes and therefore the computations per injection are dominated by step 7.

For the single layer case a similar analysis can be done, with the same asymptotic behaviour. Results are presented in

Table 2. We find that the computational costs per current injection are higher than for the double layer formalism and that this phase is dominated by step 7.

Table 1. The different phases of the Woodbury application for the double layer case.

| Phase | Line | Step | Equation | Complexity |
|---|---|---|---|---|
| Preparation | 1 | Choose $\boldsymbol{\sigma}_{old}$, $N_{inj}$ injections $\boldsymbol{j}$ | | |
| | 2 | Compute $A_{old}$ and $\boldsymbol{j}$ | | $\propto N^2$ |
| | 3 | LU decomposition $A_{old}$ | | $\frac{1}{3}N^3$ |
| | 4 | Compute $Y_{old}$ | (35) | $\frac{1}{2}N^3$ |
| | 5 | Solve $A_{old}\boldsymbol{\psi}_{old} = S\boldsymbol{j}$ | (28) | $\frac{1}{2}N^2 N_{inj}$ |
| Sigma update | 6 | Computation and LU decomposition of $I + W_\sigma^T Y_{old}$ | (37) | $\frac{1}{24}N^3$ |
| Per injection | 7 | Solve $(I + W_\sigma^T Y_{old})\boldsymbol{x}_\sigma = W_\sigma^T L_{old}\boldsymbol{\psi}_{old}$ | (37) | $\frac{1}{4}N^2$ |
| | 8 | Compute $\boldsymbol{\psi}_\sigma = L_\sigma^{-1}(\boldsymbol{\psi}_{old} - Y_{old}\boldsymbol{x}_\sigma)$ | (38) | $\frac{1}{2}N N_{elc}$ |

Table 2. The different phases of the Woodbury application for the single layer case.

| Phase | Line | Step | Equation | Complexity |
|---|---|---|---|---|
| Preparation | 1 | Choose $\boldsymbol{\sigma}_{old}$, $N_{inj}$ injections $\boldsymbol{j}$ | | |
| | 2 | Compute $A_{old}$ and $H^{-1}S$ | | $\propto N^2$ |
| | 3 | LU decomposition $A_{old}^T$ | | $\frac{1}{3}N^3$ |
| | 4 | Compute $Y_{old}$ | (35) | $\frac{1}{2}N^3$ |
| Sigma update | 5 | Computation and LU decomposition of $(I + W_\sigma^T Y_{old})^T$ | (40) | $\frac{1}{24}N^3$ |
| Per injection | 6 | Solve $(I + W_\sigma^T Y_{old})^T \boldsymbol{z}_\sigma = Y_{old}^T L_\sigma^{-1}\boldsymbol{j}$ | (40) | $\frac{1}{4}N^2$ |
| | 7 | Solve $A_{old}^T \boldsymbol{\varphi}_\sigma = L_{old} L_\sigma^{-1}\boldsymbol{j} - W_\sigma \boldsymbol{z}_\sigma$ | (41) | $\frac{1}{2}N^2 N_{elc}$ |
| | 8 | Compute $\boldsymbol{\psi}_\sigma = H^{-1}S\boldsymbol{\varphi}_\sigma$ | (42) | $\frac{1}{2}N N_{elc}$ |

*Simulations*

The accuracy of four different BEM variants are compared: single layer versus double layer and piecewise constant versus piecewise linear interpolation. For the evaluation of each BEM variant the analytical solution of the concentric sphere model presented in [25] was used as ground truth. This model consists of three concentric spherical compartments with different radii and conductivities. On the electrically isolated outer sphere two circular electrodes of finite size are attached into which opposite currents are injected. The solution of this model was found by first considering the axially symmetric problem, where a current was injected through a spherical cap, and extracted through the remainder of the outer surface. Because of the axial symmetry, a solution can be derived in terms of Legendre polynomials of $\cos\vartheta$. By rotating this configuration and subtracting this rotated configuration with opposite injection, the injected current cancels, except at the two electrodes. By this, one finds a solution for arbitrarily placed electrodes. In the simulations performed in the present study an electrode size of 0.5 cm was taken in the analytical model.

The interpolation functions used in the BEM were derived from triangulated spheres covered with an even distribution of a given number of points. The points were derived by adopting a $z$-axis, and taking approximately equal steps on the sphere in $\vartheta$ and $\varphi$ directions. Because largest spatial variations of the potentials are expected on the outermost sphere, in all cases, 50% of the nodes were placed on the outer sphere, 30% on the middle sphere and 20 % on the innermost sphere. Triangulations were obtained using a spherical Delaunay triangulation of these points. Since the results of the BEM computations are somewhat dependent on the relative orientations of the $z$-axes from which the discretization points were derived, 20 random rotations of the triangulated spheres were taken and on each combination of rotated spheres the BEM procedure was executed. The averaged BEM solutions and their standard deviations are presented. For linear potential interpolation the number of nodes was set half as large as in the corresponding constant potential interpolation approach in order to keep matrix sizes comparable.

In the BEM models with piecewise constant interpolation the injected currents were constrained to the two triangles closest to the two electrode positions. In the case of linear potential interpolation, the closest vertex of the triangular grid to the injection electrodes were determined. The amplitudes of the corresponding interpolation functions were scaled such that the total injected current equalled that of the analytical model. The amplitudes of the other interpolation functions were set to zero. Although in this way the electrode size of the BEM scales with the grid size, we verified in computations with the analytical solution that the electrode size only has a very local effect on the potential and therefore it was not necessary to correct for electrode size.

In the simulated model the outer radii for the skin, skull and brain compartments were 10, 9 and 8.5 cm, and the conductivities were set to 0.0032, 0.000049 and 0.0032 (Ωcm)$^{-1}$. A realistic electrode grid was assumed consisting of 84 electrodes evenly distributed over the skin. Current was injected through the vertex electrode Cz and extracted in turn from each of the other electrodes, when separated more than 6 cm from Cz. To find out if the BEM accuracy systematically depends on the true skull conductivity we also compute the BEM error as function of skull conductivity.

Potential distributions generated by each pair were computed with respect to average reference and compared using the following error measures

$$ADM = \frac{1}{N_{pair}}\sum_{k=0}^{N_{pair}-1}\sqrt{\frac{1}{N_{elc}}\sum_{i=0}^{N_{elc}-1}\left(\psi_{i,k}^{anal} - \psi_{i,k}^{BEM}\right)^2} \qquad (43)$$

$$RDM = \frac{1}{N_{pair}}\sum_{k=0}^{N_{pair}-1}\sqrt{\frac{1}{N_{elc}}\sum_{i=0}^{N_{elc}-1}\left(\frac{\psi_{i,k}^{anal}}{\|\psi_k^{anal}\|} - \frac{\psi_{i,k}^{BEM}}{\|\psi_k^{BEM}\|}\right)^2} \qquad (44)$$

Here $\psi_{i,k}^{BEM}$ is the BEM potential at electrode $i$, caused by a current injection applied to electrode pair $k$. ADM is an absolute difference measure and RDM is a relative difference measure, where amplitude effects are divided out using $\|\psi_k^{anal}\| \equiv \sqrt{\sum_i\left(\psi_{i,k}^{anal}\right)^2}$. The number of injected currents is denoted by $N_{pair}$. The potentials at the injection and extraction electrodes themselves were ignored in these formulas.

We also determined the gain in speed when the Woodbury formula was used, compared to direct computations. This comparison was differentiated into the *initialization phase* and the *update phase* as defined in Table 1 and

Table 2. In the direct approach the initialization phase includes the computation of the matrix elements of $A_\sigma$ and its LU decomposition. In the update phase $A_\sigma$ is updated according to (25) and the LU decomposition of the modified $A_\sigma$ is computed. Computations were performed on a Windows PC with 8 Gb memory, and 4 cores at 3201 MHz.

### Results

In Figure 2 a comparison of the accuracies of four different BEM variants is shown. Left panel shows the RDM as function of the number of discretization points whereas the right panel presents the ADM. Both panels yield the same conclusion that most accurate results are obtained when linear interpolation is combined with the double layer formalism. Remarkably, for piecewise constant interpolation results for single and double layer formulation are almost identical. Since the seeds used in the randomization of the meshes were different for the single and double layer simulations, the observed similarities of both approaches represents an averaging effect. The standard deviations, computed over different mesh orientations are quite large, indicating that the accuracy is dependent on the relative orientation of the meshes.

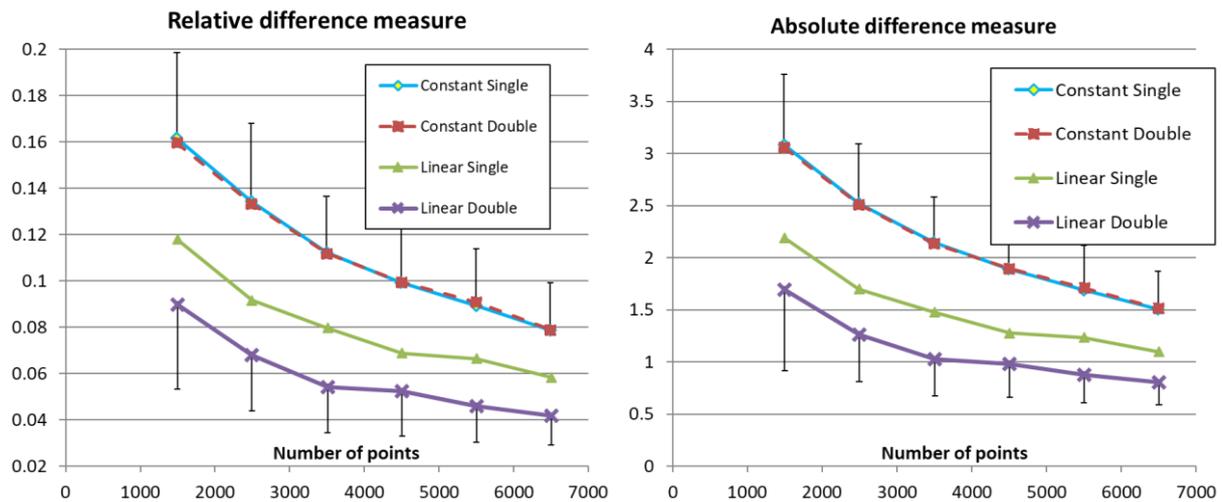

*Figure 2. The relative and absolute difference measures for four different BEM variants are shown. The standard deviation, computed over 20 realizations of random grid orientation is represented for the double layer formalism. The standard deviations of the other simulation are not shown in order to avoid clutter. These values are similar to the ones that are shown.*

In Figure 3 the RDM in the BEM is computed for a large range of skull conductivity values. In the simulations, the skin and brain conductivity were taken equal and the skin to skull conductivity ratio was varied. The matrix size of $A_\sigma$ was set equal to 5500. Again, one observes that the linear BEM combined with the double layer formalism gives the smallest errors. One also observes that for all BEM variants explored, there is some dependence on skull conductivity. For the linear BEM combined with double layer, there is a 50% increase in error when the conductivity ratio increases from 0.01 to 0.1.

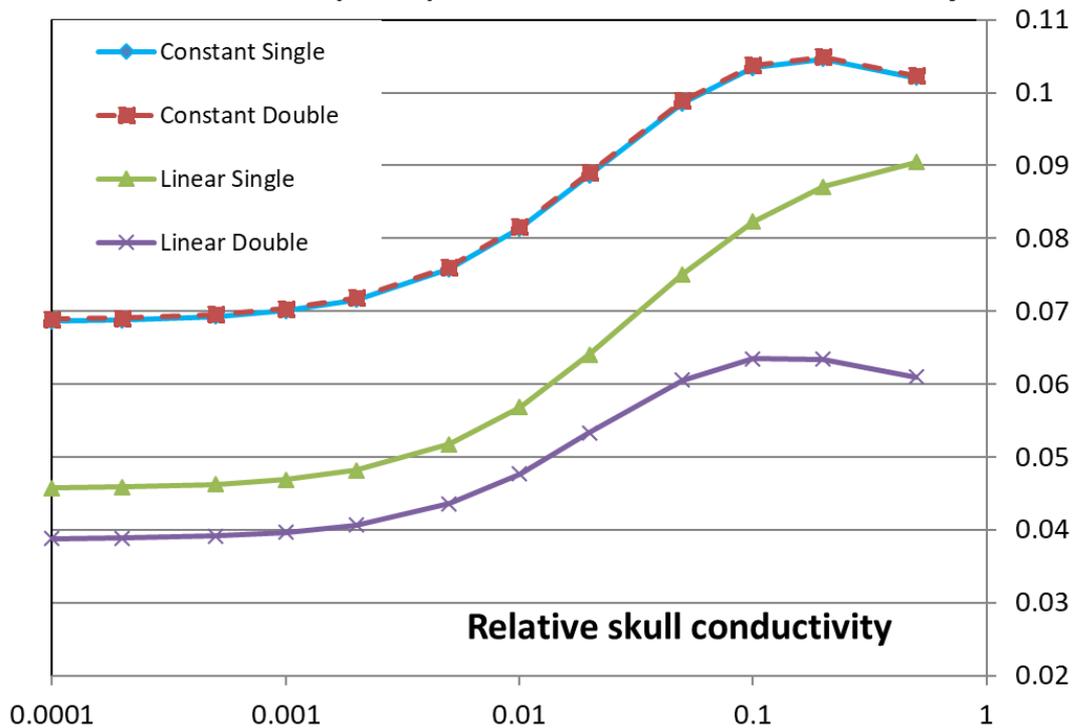

*Figure 3. The relative BEM error is plotted as function of relative skull conductivity for different BEM variants. The conductivity of brain and skin are taken equal, the skull to skin conductivity ratio is varied.*

Figure 4 shows the initialization times of the BEM for different BEM variants in seconds. The left panel represents BEM variants with piecewise constant interpolation, the right panel piecewise linear interpolation. The broken lines give the computation times for the matrix elements and LU decompositions of $A_\sigma$. In the case of single layer BEM, the computation of the required rows of $S$ is also included in the represented computation times. The continuous lines include the extra work needed to prepare for application of the Woodbury update rule. One observes that the computational costs of the initialization phase are much larger for the case of linear interpolation than for constant interpolation, indicating that these costs are dominated by the computation of the matrix elements and not by the LU decomposition, which only depends on the matrix size. The initialization costs for the single and double layer formalism are almost identical when constant interpolation is used, which can be explained by the fact that the extra computation of $S$ in the single layer formalism only concerns a few rows, corresponding to the electrode positions. One observes that for all BEM variants the extra costs to prepare for the Woodbury approach are modest (in all cases less than 50% of the total).

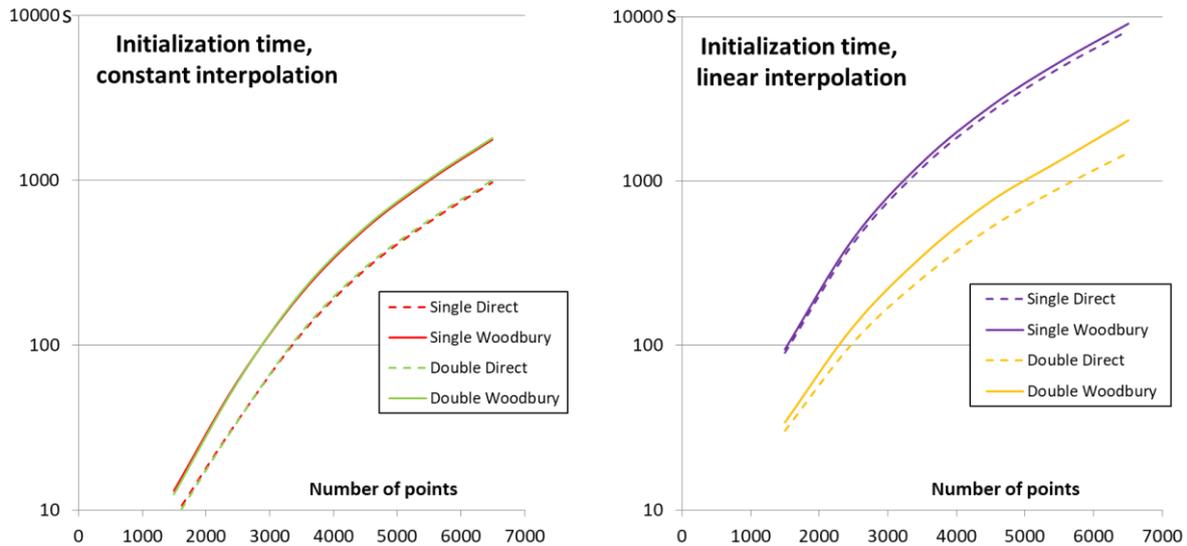

*Figure 4. The initialization time [seconds] is plotted for several BEM variants, in the left panel for piecewise constant interpolation and in the right panel for piecewise linear interpolation. The dotted lines represent the direct computations (matrix elements and LU decomposition) whereas the continuous lines represent the additional time needed to prepare for the application of the Woodbury formula.*

Finally, in Figure 5 we depict the gain in speed when the Woodbury approach is used for a new combination of $\sigma$, as opposed to the direct calculation based on equation (25). It can be observed that for all BEM combinations the gain in speed ranges from a factor of 20 to 30, for matrix sizes of 3500 and larger.

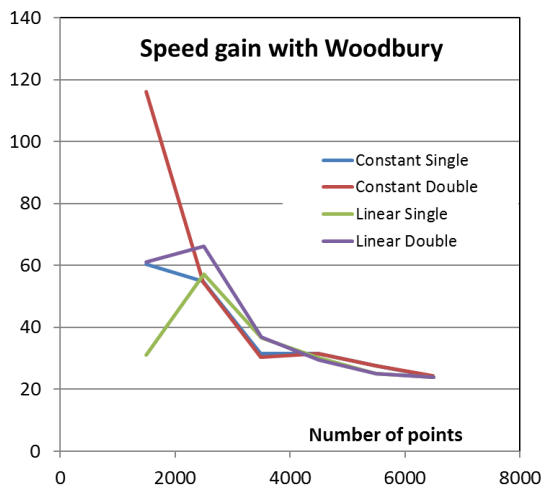

*Figure 5. The gain in speed is plotted when the Woodbury formula is used, compared to the direct computations.*

## Discussion

Our main result is that the proposed $\sigma$ update scheme is applicable for both single and double layer BEM with Galerkin discretization. This is an important extension of earlier results by Gonçalves et al [20], who treated the collocation discretization, because the only convergence results of the BEM that have been presented in the literature are based on the Galerkin with the weak formulation. For the collocation approach it is not known whether the numerical solution of the BEM in 3D converges to the true solution when the triangles become infinitely small [38]. The gain in speed achieved with this update scheme is at least a factor of 20, whereas only 50 % extra time is required in the preparation phase of the algorithm. The complexity analysis presented in Table 1 and Table 2 indicated that for $N \to \infty$ the gain in speed is reduced to a factor of 8 (equivalent to the LU decomposition of half the matrix size), which is still substantial. Application of this scheme becomes already relevant after just a few iterations of a non-linear conductivity fitting algorithm because the total gain in speed of our proposed algorithm is then dominated by the speed of the updates.

As shown in Figure 4, in the initialization phase (computation of matrix elements and LU decomposition) the computation time is much larger for linear interpolation than for constant interpolation. This finding is somewhat unexpected because the computation of the matrix elements has a complexity of $\mathscr{O}(N^2)$ compared to LU that has $\mathscr{O}(N^3)$. Therefore, for large $N$ the costs of the initialization phase is dominated by LU decomposition. This implies that within the range of $N$ explored in this study, the computation time is still dominated by the computation of the matrix elements and a substantial reduction of computational costs can be achieved by making the computation of the matrix elements more efficient, e.g., by simplifying the numerical integration for triangles and viewpoints that are relatively far apart.

Our results presented in Figure 2 show that of all BEM variants that were studied in this paper, the combination of linear interpolation and double layer integral equations gives the most accurate results. Given that [32] have found that the single layer formalism applied to superficial dipoles was more accurate than the double layer formalism, this finding is unexpected because one would intuitively expect that currents injected at the outer surface act as similar source terms as current dipoles close to the brain surface.

One may also consider this comparison from a theoretical point of view. One of the four Calderon identities [33,34], which in our symbols implies that $\hat{S}\hat{D}^* = \hat{D}\hat{S}$. This identity has been derived for the case of a single surface, but if it would be possible to extend its validity would be extended to multiple surfaces one might not expect very much difference between the single and double layer formalisms. After all, according to (20), in the discretized system one either computes $\boldsymbol{\psi}^{single} = H^{-1}SA_{\boldsymbol{\sigma}}^{-T}H\boldsymbol{j}$ or $\boldsymbol{\psi}^{double} = A_{\boldsymbol{\sigma}}^{-1}S\boldsymbol{j}$. For constant interpolation, $H$ is a diagonal matrix, with triangle sizes on the diagonal. If the differences in triangle sizes are neglected, one has $\boldsymbol{\psi}^{single} \approx SA_{\boldsymbol{\sigma}}^{-T}\boldsymbol{j}$ and the applicability of the Calderon identity would translate to the discretized system with multiple compartments, one would find $\boldsymbol{\psi}^{single} \approx \boldsymbol{\psi}^{double}$. In our simulation, we found almost identical numerical errors when comparing the single and double layer formalisms.

Apart from the EIT approach, it has also been proposed to estimate conductivity simultaneously with dipole parameters on the basis of pure EEG data [39-41] or combined EEG/MEG data [42-44]. In the former case, either one of the conductivity parameters or the dipole strengths has to be kept at a fixed value to avoid unidentifiability of the fitted model. Such approaches also require the computation of potential distributions for many different combinations of $\boldsymbol{\sigma}$ and they could benefit from the here proposed Woodbury update formula. However, it is expected that the computational benefits are inferior compared to the EIT case, because for a given total number of nodes $N$, a smaller portion of nodes has to be placed on the outer surface for maximum accuracy and therefore the difference in rank in equation (30) will be smaller than ½.

One point of concern that may hamper routine application of the proposed procedure for calibrated head models is that the numerical error of the explored BEM variants is mildly dependent on the true conductivity of the skull itself as illustrated in Figure 3. This effect may cause an unwanted bias in the estimated conductivity. It remains to be seen in future studies whether this is really an issue. However, this is not a specific disadvantage of our proposed method. So far, this aspect has not been addressed in other studies either.

The conductivity update rule presented in this paper is quite specific for the single layer and double layer formalism. The numerically preferred symmetric BEM [32], at least preferably for dipole sources, does not seem to have a structure that enables a similar algorithm because the way the conductivity parameters appear in the system matrix is much more involved than in the single and double layer BEM. Therefore, in the conductivity update phase, the single and double layer BEM allow much finer discretization before the same time is spent as with the symmetric BEM, when implemented without efficient conductivity update mechanism. It remains this to be seen where the cut-off point is and which approach is most advantageous to achieve calibrated head models from EIT data. The same is true for more modern matrix free BEM variants, such as fast multipole [45] or panel clustering [46]. In this context we would like to point to the an alternative approach to the central topic of this paper, as very recently proposed by Maksymenko et al [26], where fast conductivity updates are achieved for a wide range of methods to solve the forward problem, including BEM and FEM.

## Acknowledgements

The research leading to these results has received funding rom the European Research Council (ERC) under grant agreement no. 767235 with name "An EEG calibration toolkit for monitoring rehabilitation of stroke patients." (MRI4DEEG), associated with the ERC grant no. 291339 (4D-EEG).

## Appendix A. Derivation of boundary integral equations

To derive the boundary integral equations we follow the derivation of [32] with some modifications to allow for a polyhedral shape of the conductor with sharp edges and corners. Also, in our case potentials are generated by externally applied currents, as opposed to internal dipole sources in [32]. For the single layer formalism, the unknown potential is generated by monolayers at the conductor interfaces. The amplitudes $\varphi(\mathbf{y})$ of these layers are adjusted to comply with the interface conditions between compartments and the Neumann condition at the outer surface,

$$\psi(\mathbf{y}) = \lim_{\varepsilon \to 0} \sum_k \oiint_{\Gamma_k \setminus \partial B(\mathbf{y},\varepsilon)} \varphi(\mathbf{x}) G(\mathbf{x}-\mathbf{y}) dS_x \tag{A.1}$$

Here $G(\mathbf{x}-\mathbf{y}) = \frac{1}{4\pi}\frac{1}{|\mathbf{x}-\mathbf{y}|}$. Because $\psi(\mathbf{y})$ is a potential generated by monolayers, $\psi(\mathbf{y})$ is continuous when $\mathbf{y}$ crosses a surface $\Gamma_k$, whereas its normal derivative $\partial_n \psi = \frac{\partial \psi(\mathbf{y})}{\partial \mathbf{n}}$ makes a jump [33,34]. When $\mathbf{y} \in \Gamma_i$ one has

$$\partial_n \psi_i^{\pm} = \mp \frac{\Omega^{\pm}}{4\pi} \varphi_i + \sum_k \hat{D}_{ik}^* \varphi_k \tag{A.2}$$

Here $\Omega^{\pm}(\mathbf{y})$ are the inner (-) and outer (+) solid angles of the surface at $\mathbf{y}$ and $\hat{D}_{ik}^*$ is the adjoint dipole layer operator introduced in equation (5). For the difference in normal current $\sigma \partial_n \psi$ at both sides of the compartment interfaces one finds, using (A.2)

$$\sigma_i^+ \partial_n \psi_i^+ - \sigma_i^- \partial_n \psi_i^- = -\frac{\sigma_i^+ \Omega_i^+ + \sigma_i^- \Omega_i^-}{4\pi} \varphi_i + (\sigma_i^+ - \sigma_i^-) \sum_k \hat{D}_{ik}^* \varphi_k \tag{A.3}$$

Since on all interfaces these differences vanish, except on $\Gamma_0$ where a current $j_0(\mathbf{x})$ is injected, one finds for the monopolar distribution

$$\frac{\sigma_i^+ \Omega_i^+ + \sigma_i^- \Omega_i^-}{4\pi} \varphi_i - (\sigma_i^+ - \sigma_i^-) \sum_k \hat{D}_{ik}^* \varphi_k = \delta_{i0} j_0 \tag{A.4}$$

After solving (A.4) one finds the potential distribution with (A.1), or expressed in terms of the integral operator notation,

$$\psi_i = \sum_k \hat{S}_{ik} \varphi_k \tag{A.5}$$

Here $\hat{S}_{ik}$ is the boundary integral operator defined in (3).

In the double layer formalism, one starts with a (modified) dipolar representation

$$\sigma(\mathbf{y})\psi(\mathbf{y}) = \lim_{\varepsilon \to 0} \sum_k \oiint_{\Gamma_k \setminus \partial B(\mathbf{y},\varepsilon)} \chi(\mathbf{x}) \frac{\partial G(\mathbf{x}-\mathbf{y})}{\partial \mathbf{x}} \cdot \mathbf{n}(\mathbf{x}) dS_x + \oiint_{\Gamma_0} j_0(\mathbf{x}) G(\mathbf{x}-\mathbf{y}) dS_x \tag{A.6}$$

The modifications imply that not $\psi$ rather than $\sigma\psi$ is represented and that a monopolar source term is added. Because of the dipole layer, $\sigma\psi$ will make jumps $\chi_i$ when $\mathbf{y}$ crosses a boundary:

$$\sigma_i^+ \psi_i^+ - \sigma_i^- \psi_i^- = \chi_i \tag{A.7}$$

Note that the source term of (A.6), being a monopole potential does not contribute to the jump. Because of its construction,

$$\sigma_i^+ \partial_n \psi_i^+ - \sigma_i^- \partial_n \psi_i^- = 0 \tag{A.8}$$

and

$$\sigma_0^- \partial_n \psi_0^- = j_0 \tag{A.9}$$

Therefore, this dipole layer representation automatically satisfies the continuity conditions of normal current and the Neumann boundary condition. When also the continuity of the potential is enforced, $\psi_i^+ - \psi_i^- = 0$, one finds

$$\chi_i = (\sigma_i^+ - \sigma_i^-)\psi_i^- = (\sigma_i^+ - \sigma_i^-)\psi_i^+ \tag{A.10}$$

Hence, from now on the + and − sign can be dropped from $\psi_i^+$. The values of the double layer potential on the inner and outer sides of the boundaries are derived in [33,34]. Applied to $\sigma\psi$ and presented in our symbols one has

$$\sigma_i^{\pm}\psi_i = \mp\frac{\Omega^{\mp}}{4\pi}\chi_i + \sum_k \widehat{D}_{ik}\chi_k + \widehat{S}_{i0}j_0 \tag{A.11}$$

Substituting (A.11) in (A.10) and using that $\Omega^+ + \Omega^- = 4\pi$ one finds the well-known integral equation for the potential

$$\frac{\sigma_i^+\Omega_i^+ + \sigma_i^-\Omega_i^-}{4\pi}\psi_i - \sum_k(\sigma_k^+ - \sigma_k^-)\widehat{D}_{ik}\psi_k = \widehat{S}_{i0}j_0 \tag{A.12}$$

## Appendix B. BEM matrix elements

In this appendix some computational details are given of the computation of the matrix elements as given in equations (14) to (16). It is here assumed that the interpolation functions are derived from triangular meshes representing the compartment interfaces. To facilitate notation it will here be assumed that there is only one surface and therefore the lower indices $i$ and $k$ will be dropped. The piecewise constant case is treated in several other papers but for completeness it is treated also in this appendix. For the piecewise linear case a few novel results are presented. Matrix elements are computed partly analytically and partly numerically. For numerical integration a weighted sum over 16 points inside the triangle was used [47].

*Piecewise constant case*

In the piecewise constant case the support of $h^n(\mathbf{x})$ extends to a single triangle and the numbering of the interpolation functions corresponds to the numbering of the triangles. We have

$$(H)^{m,n} = \langle h^m, h^n \rangle = A^n \delta_{m,n} \tag{B.1}$$

where $A^n$ is the area of triangle $n$.

The computation of the matrix elements $(G)^{m,n}$ and $(S)^{m,n}$ consist of two nested integrals. The outer integrals over the view point $\mathbf{y}$ in (14) and (16) must be computed numerically and here we focus on $\widehat{D}h^n$ and $\widehat{S}h^n$. For $\widehat{D}h^n$ computes (apart from the factor $\frac{1}{4\pi}$)

$$\Omega(\mathbf{x}_0, \mathbf{x}_1, \mathbf{x}_2) \equiv \iint_\Delta \nabla R^{-1} \cdot \mathbf{n}\, dS_x = 2\operatorname{atan}\frac{d}{|x_0||x_1||x_2| + |x_0|(x_1 \cdot x_2) + |x_1|(x_2 \cdot x_0) + |x_2|(x_0 \cdot x_1)} \tag{B.2}$$

Here the view point has been shifted to the origin, $\mathbf{x}_0$, $\mathbf{x}_1$ and $\mathbf{x}_2$ denote the shifted triangle corners and $d \equiv \mathbf{x}_0 \cdot (\mathbf{x}_1 \times \mathbf{x}_2)$ is the determinant and $R = |\mathbf{x}|$. Note that there is no singularity for the case that $d = 0$, except when one of the corners coincides with the origin. Equation (B.2) has been presented in [48] and derived from first principles in [49]. For $\widehat{S}h^n$ one computes (again skipping the factor $\frac{1}{4\pi}$)

$$Y = \iint_\Delta R^{-1} dS = -\frac{d}{A}\Omega(\mathbf{x}_0, \mathbf{x}_1, \mathbf{x}_2) + \frac{\mathbf{n}}{A} \cdot \sum_k (\mathbf{x}_k \times \mathbf{x}_{k+1})L_k^{(-3)} \tag{B.3}$$

with

$$L_k^{(-3)} \equiv \frac{1}{|x_{k+1} - x_k|}\log\frac{(x_{k+1}-x_k)\cdot x_k + |x_{k+1}-x_k||x_k|}{(x_{k+1}-x_k)\cdot x_{k+1} + |x_{k+1}-x_k||x_{k+1}|} \tag{B.4}$$

and $\mathbf{n} \equiv \mathbf{x}_0 \times \mathbf{x}_1 + \mathbf{x}_1 \times \mathbf{x}_2 + \mathbf{x}_2 \times \mathbf{x}_0$ and $A \equiv |\mathbf{n}|$ is twice the triangle area. This equation has been presented in [27], but with printing errors. The logarithm in (B.4) is singular when one of the corner points is zero. This situation does not happen in the Galerkin approach of the BEM when the sample points of the numerical integration are located inside the triangles. However, with the collocation approach, using linear interpolation and the electrode size extended over several triangles, one has to deal with $\mathbf{x}_k = 0$. Assuming $k = 0$ one can compute $Y$ from scratch and parameterize $\Delta$ as $\mathbf{x} = v(\mathbf{x}_1 + u(\mathbf{x}_2 - \mathbf{x}_1))$ with $(u,v) \in [0,1] \times [0,1]$. Then $\frac{\partial \mathbf{x}}{\partial u} \times \frac{\partial \mathbf{x}}{\partial v} = v(\mathbf{x}_2 \times \mathbf{x}_1)$ and hence one finds

$$Y = \int_0^1 \int_0^1 \frac{v}{v|\mathbf{x}_1+u(\mathbf{x}_2-\mathbf{x}_1)|} du\, dv = \frac{1}{|x_2-x_1|} \log \frac{(\mathbf{x}_2-\mathbf{x}_1)\cdot\mathbf{x}_1+|x_2-x_1||x_1|}{(\mathbf{x}_2-\mathbf{x}_1)\cdot\mathbf{x}_2+|x_2-x_1||x_2|} = L_1^{(-3)} \quad (B.5)$$

*Piecewise linear case*

In the case of piecewise linear interpolation, the numbering of the interpolation functions corresponds to the vertices of the triangulated surface. For the computation of the $(n,m)$ matrix elements one has to distinguish three cases: complete overlap ($n=m$), semi overlap (vertex $n$ is neighbor of vertex $m$) and no overlap. In the first case the integrals consist of a sum of integrals over all triangles adjacent to vertex $n$. These triangles are denoted by $\Delta_n^j$ where $j$ runs over the set of adjacent triangles $T_n$. In the semi overlap case the integrals consist of a sum of two triangle integrals, at both sides of the edge $(n,m)$. These triangles are denoted by $\Delta_{nm}^1$ and $\Delta_{nm}^2$.

For the matrix elements $H^{m,n}$ one has

$$(H)^{m,n} = \begin{cases} \frac{1}{12}\sum_{j\in T_n} A(\Delta_n^j) & m = n \\ \frac{1}{12}\left(A(\Delta_{nm}^1) + A(\Delta_{nm}^2)\right) & m \text{ neigbor of } n \\ 0 & \text{otherwise} \end{cases} \quad (B.6)$$

where $A(\Delta)$ denotes the area of triangle $\Delta$.

The matrix elements $(G)^{m,n}$ and $(S)^{m,n}$ are more involved. When (14) is reduced to one surface the definition of $(G)^{m,n}$ is

$$(G)^{m,n} = \langle h^m, \widehat{D}h_k^n \rangle + \langle h^m, \frac{\Omega^-}{4\pi} h^n \rangle \quad (B.7)$$

We first consider the case of no overlap, for which the second term is zero.

The formulas for the piecewise linear case of $\widehat{D}h^n$ have been derived in [27]. We here show that with a small generalization of this approach, similar formulas can be found for $\widehat{S}h^n$. For both $\widehat{D}h^n$ and $\widehat{S}h^n$ we have linearly varying functions over a triangle and the following integrals are needed, for different values of $t$

$$\Gamma_k^{(t)} \equiv \frac{1}{d}\iint_\Delta R^t\, \mathbf{z}_k \cdot \mathbf{x}\, dS \quad k \in \{0,1,2\}, t \in \{-1,-3\} \quad (B.8)$$

where $\mathbf{z}_k \equiv \mathbf{x}_{k-1} \times \mathbf{x}_{k+1}$ (with $\mathbf{x}_{-1} \equiv \mathbf{x}_2$ and $\mathbf{x}_3 \equiv \mathbf{x}_0$). The computation of $\widehat{S}h^n$ corresponds to the case $t=-1$ and $\widehat{D}h^n$ corresponds to $t=-3$ because the linearly weighted solid angles are given by

$$\frac{1}{d}\iint_\Delta (\mathbf{z}_k \cdot \mathbf{x})\nabla R^{-1} \cdot \mathbf{n} dS = -\frac{1}{A}\iint_\Delta (\mathbf{z}_k \cdot \mathbf{x})R^{-3} dS = -\frac{1}{A}\Gamma_k^{(-3)} \quad (B.9)$$

An analytical expression for (B.8) is derived by considering the following integral $\boldsymbol{\Gamma}^{(t)}$ and computing it in two different manners. The first manner is to use Stokes' theorem,

$$\boldsymbol{\Gamma}^{(t)} \equiv \iint_\Delta \nabla R^{t+2} \times \mathbf{n} dS = \iint_\Delta \nabla \times R^{t+2}\mathbf{n} dS = \oint_\Delta R^{t+2} d\boldsymbol{\gamma}$$

$$= \sum_{k=0,1,2} \int_{\mathbf{x}_k}^{\mathbf{x}_{k+1}} R^{t+2} d\boldsymbol{\gamma} = \sum_{k=0,1,2} \left(L_{k-1}^{(t)} - L_k^{(t)}\right)\mathbf{x}_k \quad (B.10)$$

where

$$L_k^{(t)} \equiv \int_0^1 \|\mathbf{y}_k + \gamma(\mathbf{y}_{k+1} - \mathbf{y}_k)\|^{t+2} d\gamma \quad (B.11)$$

These line integrals can be computed analytically. The case $t=-3$ is given by (B.4) and for $t=-1$ one finds after some algebra

$$L_k^{(-1)} = \frac{(\mathbf{x}_{k+1}-\mathbf{x}_k)\cdot\mathbf{x}_{k+1}|\mathbf{x}_{k+1}|-(\mathbf{x}_{k+1}-\mathbf{x}_k)\cdot\mathbf{x}_k|\mathbf{x}_k|}{2|\mathbf{x}_{k+1}-\mathbf{x}_k|^2} + \frac{|\mathbf{x}_{k+1}\times\mathbf{x}_k|^2}{2|\mathbf{x}_{k+1}-\mathbf{x}_k|^3} \log \frac{(\mathbf{x}_{k+1}-\mathbf{x}_k)\cdot\mathbf{x}_k+|\mathbf{x}_{k+1}-\mathbf{x}_k||\mathbf{x}_k|}{(\mathbf{x}_{k+1}-\mathbf{x}_k)\cdot\mathbf{x}_{k+1}+|\mathbf{x}_{k+1}-\mathbf{x}_k||\mathbf{x}_{k+1}|} \quad (B.12)$$

The alternative manner to work out $\boldsymbol{\Gamma}^{(t)}$ is to carry out the differentiation of $R^{t+2}$ and move $\mathbf{n}$ in front of the integral

$$\boldsymbol{\Gamma}^{(t)} = \iint_\Delta (t+2)R^t\ \boldsymbol{x}\times\boldsymbol{n}\mathrm{d}S = -\frac{t+2}{A}\boldsymbol{n}\times\iint_\Delta R^t\ \boldsymbol{x}\,\mathrm{d}S \qquad (B.13)$$

Using the identity $\boldsymbol{x} = \frac{1}{d}\sum_{k=0,1,2}(\boldsymbol{z}_k\cdot\boldsymbol{x})\boldsymbol{x}_k$ one finds an expression wherein the integrals of interest $\Gamma_k^{(t)}$ appear

$$\boldsymbol{\Gamma}^{(t)} = -\frac{t+2}{Ad}\boldsymbol{n}\times\sum_{k=0,1,2}\boldsymbol{x}_k\iint_\Delta R^t\ \boldsymbol{z}_k\cdot\boldsymbol{x}\,\mathrm{d}S = -\frac{t+2}{A}\boldsymbol{n}\times\sum_{k=0,1,2}\boldsymbol{x}_k\Gamma_k^{(t)} \qquad (B.14)$$

In order to enable comparison of coefficients of $\boldsymbol{x}_k$ in (B.10) and (B.14) and hence to find equations for the integrals of interest, one needs to work out $\boldsymbol{n}\times\boldsymbol{x}_k$. We do so by applying the vector identity $\boldsymbol{a}\times(\boldsymbol{b}\times\boldsymbol{c}) = \boldsymbol{b}(\boldsymbol{a}\cdot\boldsymbol{c}) - \boldsymbol{c}(\boldsymbol{a}\cdot\boldsymbol{b})$ to find

$$\boldsymbol{n}\times\boldsymbol{x}_k = \sum_{l=0,1,2}(\boldsymbol{x}_{l-1}\times\boldsymbol{x}_{l+1})\times\boldsymbol{x}_k = \sum_{l=0,1,2}((\boldsymbol{x}_{l+1}-\boldsymbol{x}_{l-1})\cdot\boldsymbol{x}_k)\boldsymbol{x}_l \qquad (B.15)$$

and arrive at

$$\boldsymbol{\Gamma}^{(t)} = \frac{t+2}{A}\sum_{k,l}((\boldsymbol{x}_{l-1}-\boldsymbol{x}_{l+1})\cdot\boldsymbol{x}_k)\Gamma_k^{(t)}\ \boldsymbol{x}_l \qquad (B.16)$$

Comparing coefficients of (B.10) and (B.16) gives the following system of equations:

$$\sum_{k=0,1,2}((\boldsymbol{x}_{l-1}-\boldsymbol{x}_{l+1})\cdot\boldsymbol{x}_k)\,\Gamma_k^{(t)} = \frac{A}{t+2}\left(L_{l-1}^{(t)} - L_l^{(t)}\right) \quad l\in\{0,1,2\} \qquad (B.17)$$

Because these equations add up to zero they are linearly dependent and the last one is replaced by the following equation: $\Gamma_0^{(t)} + \Gamma_1^{(t)} + \Gamma_2^{(t)} = \iint_\Delta R^t\mathrm{d}S$, which can be computed analytically for $t=-1$ and $t=-3$. After some more algebra, one finally finds

$$\Gamma_0^{(t)} = \frac{\boldsymbol{x}_1-\boldsymbol{x}_2}{(t+2)A}\cdot\left((\boldsymbol{x}_1-\boldsymbol{x}_0)F_0^{(t)} + (\boldsymbol{x}_2-\boldsymbol{x}_1)F_1^{(t)} + (\boldsymbol{x}_0-\boldsymbol{x}_2)F_2^{(t)}\right) + \frac{\boldsymbol{n}\cdot(\boldsymbol{x}_2\times\boldsymbol{x}_1)}{A^2}\iint_\Delta R^t\mathrm{d}S$$

$$\Gamma_1^{(t)} = \frac{\boldsymbol{x}_2-\boldsymbol{x}_0}{(t+2)A}\cdot\left((\boldsymbol{x}_1-\boldsymbol{x}_0)F_0^{(t)} + (\boldsymbol{x}_2-\boldsymbol{x}_1)F_1^{(t)} + (\boldsymbol{x}_0-\boldsymbol{x}_2)F_2^{(t)}\right) + \frac{\boldsymbol{n}\cdot(\boldsymbol{x}_0\times\boldsymbol{x}_2)}{A^2}\iint_\Delta R^t\mathrm{d}S \qquad (B.18)$$

$$\Gamma_2^{(t)} = \frac{\boldsymbol{x}_0-\boldsymbol{x}_1}{(t+2)A}\cdot\left((\boldsymbol{x}_1-\boldsymbol{x}_0)F_0^{(t)} + (\boldsymbol{x}_2-\boldsymbol{x}_1)F_1^{(t)} + (\boldsymbol{x}_0-\boldsymbol{x}_2)F_2^{(t)}\right) + \frac{\boldsymbol{n}\cdot(\boldsymbol{x}_1\times\boldsymbol{x}_0)}{A^2}\iint_\Delta R^t\mathrm{d}S$$

For $t=-1$ one uses $\iint_\Delta R^{-1}\mathrm{d}S = Y$, which is worked out in (B.3) and (B.5). For $t=-3$ one has

$$\iint_\Delta R^{-3}\mathrm{d}S = -\frac{A}{d}\iint_\Delta \nabla R^{-1}\cdot\boldsymbol{n}\mathrm{d}S = -\frac{A}{d}\Omega(\boldsymbol{x}_0,\boldsymbol{x}_1,\boldsymbol{x}_2) \qquad (B.19)$$

which can be computed with (B.2).

Analogous to $L_k^{(-3)}$ the use of $L_k^{(-1)}$ in the computation of $(S)^{m,n}$ would be problematic when $\boldsymbol{x}_k = 0$. Again, with the Galerkin approach that situation does not occur as long as in the numerical integration of (B.17) in $\langle,\rangle$ the integration points are chosen inside the triangles. For $(G)^{m,n}$ special treatment is needed when the determinant $d=0$, which happens when $m=n$ or when $m$ and $n$ are neighbors. The diagonal element is (apart from the integration in $\langle,\rangle$)

$$\left((\widehat{D} + \frac{\Omega^-}{4\pi})h^n\right)(\boldsymbol{y}) = \frac{1}{4\pi}\left(\iint_{\cup_j \Delta_n^j}^* h^n(\boldsymbol{x})\nabla_x|\boldsymbol{x}-\boldsymbol{y}|^{-1}\cdot\boldsymbol{n}(\boldsymbol{x})\mathrm{d}S_x + \Omega^-(\boldsymbol{y})h^n(\boldsymbol{y})\right) \qquad (B.20)$$

Note that $\boldsymbol{y}$ is at one of triangles $\Delta_n^j$ adjacent to vertex $n$ and here we have $(\boldsymbol{x}-\boldsymbol{y})\cdot\boldsymbol{n}=0$. Therefore the only remaining part of the integral $\iint^*$ is the part over the spherical cap, which equals $-\Omega^-(\boldsymbol{y})h^n(\boldsymbol{y})$ and thus we have that $(G)^{n,n} = 0$. This result is analogous to [27], where the collocation approach was considered and matrix elements were defined differently.

Finally, when $m$ and $n$ are neighbors, special treatment is needed for the integration over the overlapping parts of $h^n(\boldsymbol{x})$ and $h^m(\boldsymbol{x})$. In the general case, expanding (B.7) yields

$$(G)^{n,m} = \sum_{j_1,j_2}\iint_{\Delta_m^{j_1}}\left(h^m(\boldsymbol{y})\left(\iint_{\Delta_n^{j_2}} h^n(\boldsymbol{x})\nabla_x G(\boldsymbol{x}-\boldsymbol{y})\cdot\boldsymbol{n}(\boldsymbol{x})\mathrm{d}S_x + \frac{\Omega^-(\boldsymbol{y})h^n(\boldsymbol{y})}{4\pi}\right)\right)\mathrm{d}S_y \qquad (B.21)$$

On the two triangles $\Delta_{nm}^1$ and $\Delta_{nm}^2$ we have that $\nabla_x G(x - y) \cdot n(x) = 0$ and $\Omega^-(y) = 2\pi$. Therefore, the first term vanishes and the contributions of those two triangles to the matrix elements equal $\frac{1}{24}A(\Delta_{nm}^1)$ and $\frac{1}{24}A(\Delta_{nm}^2)$. The contributions of the other triangles are not hampered by zero determinant and can be computed using the formulas given by (B.18).

Remarkably, although the inner and outer solid angles $\Omega^-$ and $\Omega^+$ at sharp corners appear in many occasions in the theoretical formulas, they never have to be computed explicitly to obtain the BEM matrix elements.

## Appendix C. Singularity of the system matrix

The standard piecewise linear and piecewise constant interpolation functions have the following properties

$$h_k^n(x_i^m) = \delta_{k,i}\delta_{n,m} \tag{C.1}$$

and

$$\sum_{n=0}^{N_k-1} h_k^n(x) = 1 \quad \text{for all } x \in \Gamma_k \quad \text{and } h_n^k(x) \geq 0 \quad \text{for all } k \tag{C.2}$$

Here $N_k$ is the number of interpolation functions associated with surface $\Gamma_k$.

In this appendix it is shown that if such functions are used in combination with the Galerkin method, the system matrix has an eigenvector $e = (e_{N_0}^T, e_{N_1}^T, \ldots, e_{N_{K-1}}^T)^T$ with eigenvalue zero. Without deflation, the system matrix $\tilde{A}_\sigma$ is given by

$$(\tilde{A}_\sigma)_{i,k} \equiv \sigma_i^+ H_{i,i}\delta_{i,k} - (\sigma_k^+ - \sigma_k^-)G_{i,k} \tag{C.3}$$

The $m$-th component on $\Gamma_i$ of this matrix multiplied with $e$ gives

$$(\tilde{A}_\sigma e)_i^m = \left(\sigma_i^+ H_{i,i}e_i - \sum_k(\sigma_k^+ - \sigma_k^-)G_{i,k}e_k\right)^m \tag{C.4}$$

$$= \sigma_i^+ \sum_n\langle h_i^m, h_i^n\rangle - \sum_k(\sigma_k^+ - \sigma_k^-)\sum_n\langle h_i^m, (\hat{D}_{i,k} + \frac{\Omega_k^-}{4\pi}\delta_{i,k})h_k^n\rangle$$

$$= \sigma_i^+ \langle h_i^m, 1\rangle - \sum_k(\sigma_k^+ - \sigma_k^-)\langle h_i^m, (\hat{D}_{i,k} + \frac{\Omega_k^-}{4\pi}\delta_{k,i})1\rangle$$

To work out the second term, the definition of the inner product $\langle,\rangle_i$ and de operator $\hat{D}_{i,k}$ are used. One finds for the second term in the last line of (C.4)

$$\langle h_i^m, (\hat{D}_{i,k} + \frac{\Omega_k^-}{4\pi}\delta_{k,i})1\rangle_i = \oiint_{\Gamma_i}\left(h_i^m(y)\left(\oiint_{\Gamma_k}^* \nabla_x G(x-y)\cdot n(x)dS_x + \frac{\Omega_k^-}{4\pi}\delta_{k,i}\right)\right)dS_y \tag{C.5}$$

The inner integral in (C.5) depends on the relative position of the view point $y$, which is on $\Gamma_i$, with respect to $\Gamma_k$. One finds

$$\oiint_{\Gamma_k}^* \nabla_x G(x-y)\cdot n(x)dS_x = \begin{cases} 0 & \Gamma_i \text{ outside } \Gamma_k \\ -\Omega_k^-/(4\pi) & \Gamma_i = \Gamma_k \\ 1 & \Gamma_i \text{ inside } \Gamma_k \end{cases} \tag{C.6}$$

Substituting this result in (C.4), it is found that

$$(\tilde{A}_\sigma e)_i^m = \langle h_i^m, 1\rangle\left(\sigma_i^+ + \sum_{\{k\in\mathbb{N}|\Gamma_i \text{ inside } \Gamma_k\}}(\sigma_k^+ - \sigma_k^-)\right) \tag{C.7}$$

The "-part" first term in this sum cancels with $\sigma_i^+$ and each "-part" of the next enclosing surface cancels with the "+part" of the previous surface. In the end the total equals $\sigma_0^+ = 0$. Hence it follows that $\tilde{A}_\sigma e = \mathbf{0}$.